# Linking vascular pathways to organ anatomical structures in the Human Reference Atlas

Katherine Gustilo[1,2], Quinlan Cuddy[1,2], Avinash Boppana[3], Bruce W. Herr II[2], Katy Börner[2], Griffin M Weber[4,5*]

[1] University of Colorado Anschutz Medical Campus, Aurora, Colorado, USA
[2] Department of Intelligent Systems Engineering, Luddy School of Informatics, Computing, and Engineering, Indiana University, Bloomington, Indiana, USA
[3] Sedona Health, Inc., New York, New York, USA
[4] Department of Biomedical Informatics, Harvard Medical School, Boston, Massachusetts, USA
[5] Department of Medicine, Beth Israel Deaconess Medical Center, Boston, Massachusetts, USA

* Corresponding author
Griffin M Weber, weber@hms.harvard.edu

# Abstract

As part of the Human Reference Atlas (HRA) initiative, which brings together more than 25 international consortia to map the entire healthy adult body, we developed a multiscale Vasculature Common Coordinate Framework (VCCF). The VCCF uses vascular pathways to describe the location of organs, anatomical structures, and cells throughout the body. Here, we present the VCCF Blood Vasculature Organ Crosswalk, which links blood vessels to anatomical structures they supply or drain, and the Pathways Crosswalk, which lists relationships between blood vessels, lymphatic vessels, and nerves. Blood Vasculature Organ Crosswalk Diagrams support the authoring and review of the crosswalk tables. In machine readable files, we catalog 349 blood vessels from these diagrams, with 663 linkages to 295 anatomical structures across 20 organs/body regions and 28 functional tissue units. The VCCF crosswalk tables and diagrams have been validated algorithmically and by 11 organ experts. They have brought together experts from different domains to document existing knowledge of the human vasculature and identify gaps where future research is needed.

# Background & Summary

Hundreds of scientists from more than 25 international consortia are part of an ambitious initiative to construct a multiscale Human Reference Atlas (HRA), which maps the entire healthy adult body from organs and tissues down to cells and biomarkers[1]. To date, the HRA has cataloged more than 4,000 anatomical structures across 33 organs. The HRA is published as a knowledge graph using standard ontologies[2]. To construct this knowledge graph, teams of domain experts fill out organ-specific spreadsheets called “Anatomical Structures, Cell Types, Plus Biomarkers” (ASCT+B) tables[3]. These are linked to 3D reference objects, single cell datasets, biospecimen metadata, and other digital objects[4]. A web portal and interactive data visualizations provide access to HRA data and enable exploration of the knowledge graph[5]. Each of these are components of the HRA’s Common Coordinate Framework (CCF)[1]. Also included in the HRA’s CCF is a Vasculature Common Coordinate Framework (VCCF), which uses vascular pathways to describe the location of organs, anatomical structures, and cells throughout the body[6] and to measure the distance between cells and their nearest blood vessels[7].

There are two types of ASCT+B tables (**Figure 1**). Most organs use “nested partonomy” ASCT+B tables. Anatomical structures in these tables are hierarchical, starting with the whole organ, then major components of the organ, then smaller subparts of those components, and so on down to functional tissue units (FTUs), which we define as “the smallest tissue organization that performs a unique physiologic function and is replicated multiple times in a whole organ”[8,9]. We use the term “part_of” to describe the relationship between different levels in this hierarchy.

There are three organs—blood vasculature, lymph vasculature, and peripheral nervous system (PNS)—which use “branching pathway” ASCT+B tables. In these tables, we use the term “branches_from” to describe the relationship between levels. For example, in the blood vasculature table, the aorta branches from the left ventricle of the heart, the brachiocephalic artery branches from the aorta, and so on down to arterioles and capillaries. A separate venous hierarchy also starts at the heart and lists the branching of veins and venules.

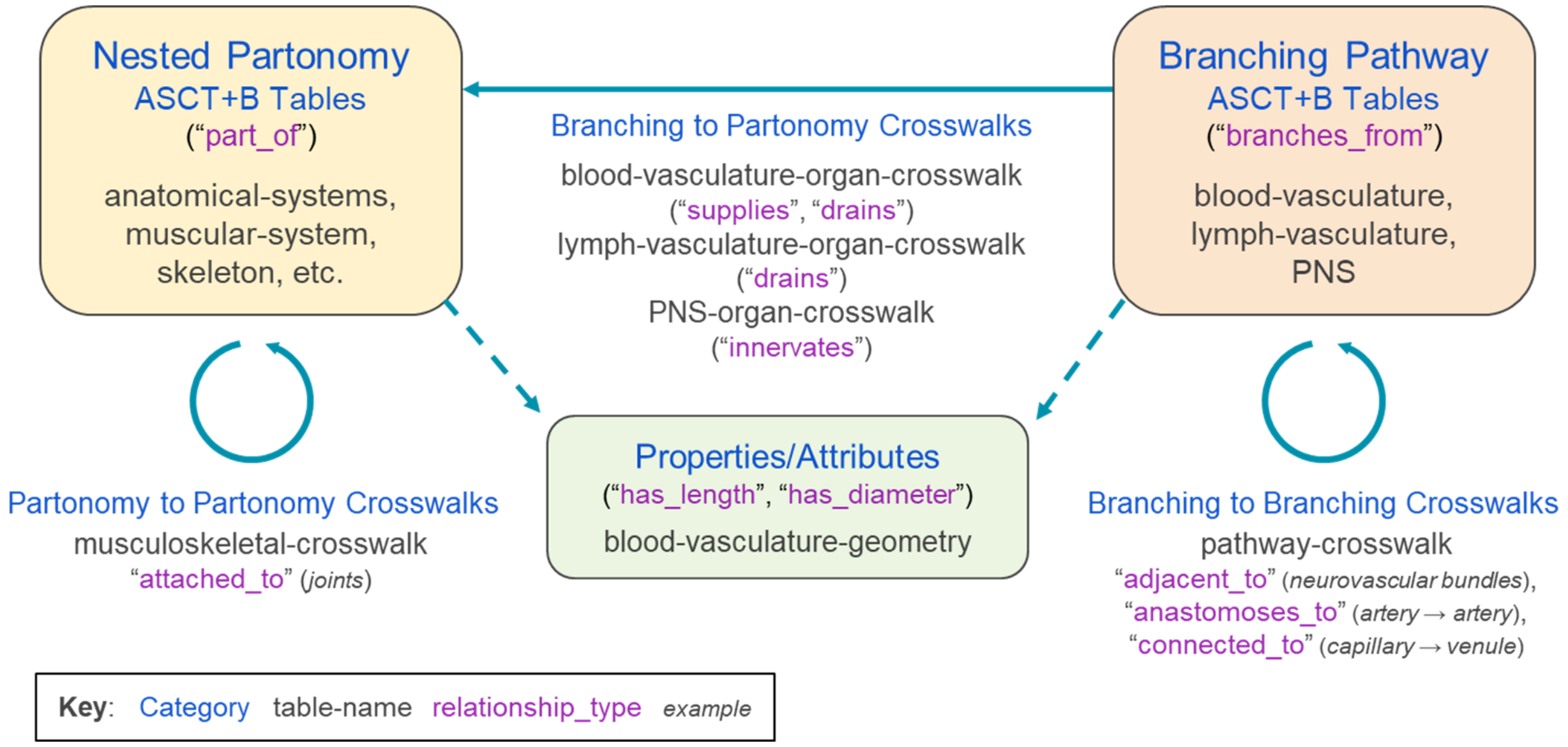

**Figure 1**. **Data tables within the Human Reference Atlas (HRA)**. Shown are two types of Anatomical Structures, Cell Types, Plus Biomarkers (ASCT+B) tables and three types of Crosswalk tables (solid blue lines). An additional Properties/Attribute table indicating geometric properties of blood vessels, such as length and diameter, is included to illustrate one of many other digital objects that are connected to ASCT+B tables in the HRA knowledge graph.

We previously described the construction and contents of this blood vasculature ASCT+B table, which includes more than 1000 blood vessels[10]. In this paper, we describe two new VCCF "crosswalk tables", which extend the HRA knowledge graph with additional relationships between blood vessels and anatomical structures in the ASCT+B tables (**Figure 1**). The first, called the "blood-vasculature-organ-crosswalk", links blood vessels to the organ anatomical structures that they supply (with oxygen, nutrients, etc.) or drain. The second, called the "pathway-crosswalk" table, indicates relationships between blood and lymph vessels and nerves, such as neurovascular bundles, blood vessel anastomoses, and connections between arterial and venous vascular trees.

In addition, we introduce blood vasculature-organ crosswalk diagrams, which depict an organ's anatomical structures in a novel way that highlights the vascular pathways through that organ. These diagrams are an essential tool in constructing and validating the crosswalks.

# Methods

Construction of the blood vasculature-organ crosswalk tables begins with domain experts (anatomists, clinicians, organ experts, etc.) developing a blood vasculature-organ crosswalk diagram (**Figure 2**), which contain two types of objects. Overlapping rectangles represent the nested partonomy of anatomical structures in an organ, from gross anatomical structures on the macro-scale (on the left) down to functional tissue units on the micro-scale (on the right). These structures come from the organ nested partonomy ASCT+B tables. Crossing through these rectangles are lines that represent blood vessels that supply, drain, or pass through those structures. The list of vessels comes from the blood vasculature branching pathway ASCT+B table.

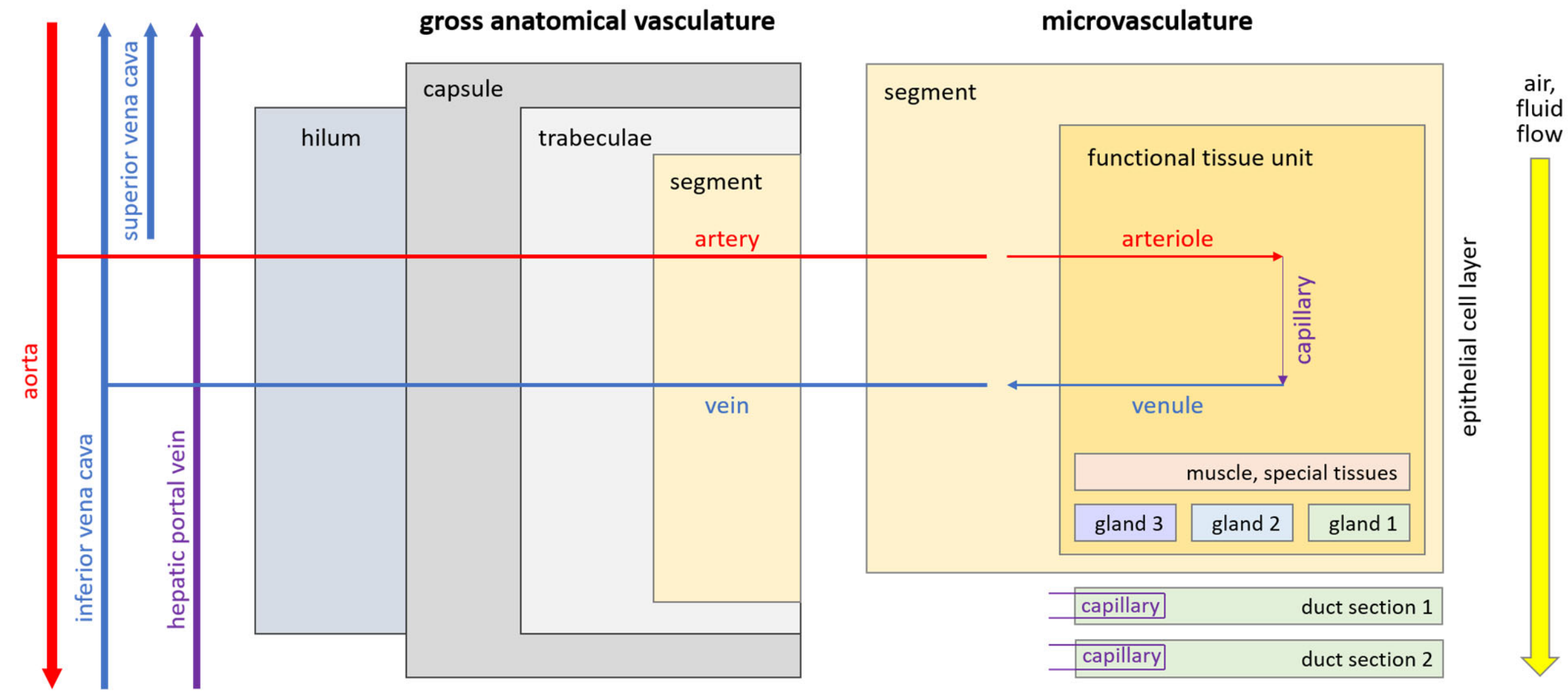

**Figure 2**. **Blood vessel-organ crosswalk diagram template**. Organ anatomical structures are represented as rectangles, overlaid by lines representing vasculature loops.

### *Placing blood vessels within the crosswalk diagram*

The diagrams are distorted, compared to the actual anatomical shape and size of the organ, to straighten the blood vessels pathways as much as possible. In constructing these diagrams, the goal is to work towards a layout that looks similar to the template in **Figure 2**, with the main blood vasculature pathway drawn as a large loop. The loop starts in the upper left with the main artery supplying the organ and continues to the right, leading to smaller arteries, then arterioles, and finally reaching a capillary, which runs downwards on the right side of the diagram. At the bottom right of the diagram, the capillary connects to venules, which lead to veins, which leave the organ in the lower left of the diagram. Blood flow is typically clockwise through the diagram, with the arrow on the capillary pointing downwards. An exception is countercurrent flow, which is counterclockwise, with the arrow on the capillary pointing upwards and veins appearing above the arteries rather than below them.

There can be more than one blood vessel loop through the organ and loops can branch off of others. Each loop corresponds to a different vasculature pattern or tissue type. However, note that there might be many (i.e., millions) of instances of a single type of blood vessel pathway in an organ, but they are drawn as a single loop in this diagram. As a result, the total number of loops is typically less than ten and approximately equal to the number of different FTU types in the organ.

### *Placing organ anatomical structures within the crosswalk diagram*

Organ anatomical structures are placed around the straightened blood vessel pathway loops. As a result, structures that are near each other in the actual 3D anatomical shape of the organ might be separated in the diagram. Usually, larger organ structures (e.g., organ capsule, lobes, etc.) will be on the left side of the diagram with the larger blood vessels. Smaller structures (e.g., organ segments, FTUs) will be on the right with capillaries. A small vertical gap typically appears near the middle of the diagram. On the left of this gap are large named structures and named blood vessels. On the right of the gap are smaller, repeated structures and blood vessels that are not individually named.

Structures with an epithelial surface are the furthest on the right. Just to the right of those structures represents the space either outside the body or the inside of a hollow organ. A large downwards arrow on the right edge of the diagram indicates the substance (e.g., fluid or air) that flows past that epithelial surface. (One can also imagine this being where the “microbiome” of the organ lives.) Deeper layers in an organ, such as the muscular layers of the intestine, are not as far to the right.

Color distinguishes different types of tissues in the organ. Warm colors indicate parenchyma and the main functional tissue of the organ (yellow as the primary color and pink for muscles or to highlight special tissues); grey indicates stroma, or connective tissue; light green indicates glands and ducts; and blue and purple are used as alternative colors for glands and special anatomical structures in certain cases for added clarity.

### *Creating the crosswalk tables*

The diagrams are then used as a visual aid to create the blood-vasculature-organ-crosswalk table and the pathway-crosswalk table by recording the relationships seen in the diagrams into spreadsheets. Without the diagrams, we found the authoring and review of these tables to be near impossible. The diagrams provide a two-dimensional representation of the complex three-dimensional vascular pathways through the organ, which proved to be essential in performing this work.

# Data Records

The blood vasculature organ crosswalk (blood-vasculature-organ-crosswalk.csv) and pathway crosswalk (pathway-crosswalk.csv) tables are available as comma-separated values (CSV) files with a Creative Commons Attribution 4.0 International (CC BY 4.0) license[11].

The blood vasculature-organ crosswalk diagrams (blood-vasculature-organ-crosswalk-diagrams.pptx) are included along with the tables as a Microsoft Powerpoint Open XML Presentation (PPTX) file and also saved to Portable Document Format (PDF), Scalable Vector Graphics (SVG), and Portable Network Graphic (PNG) formats. An example diagram for the pancreas is shown in **Figure 3**.

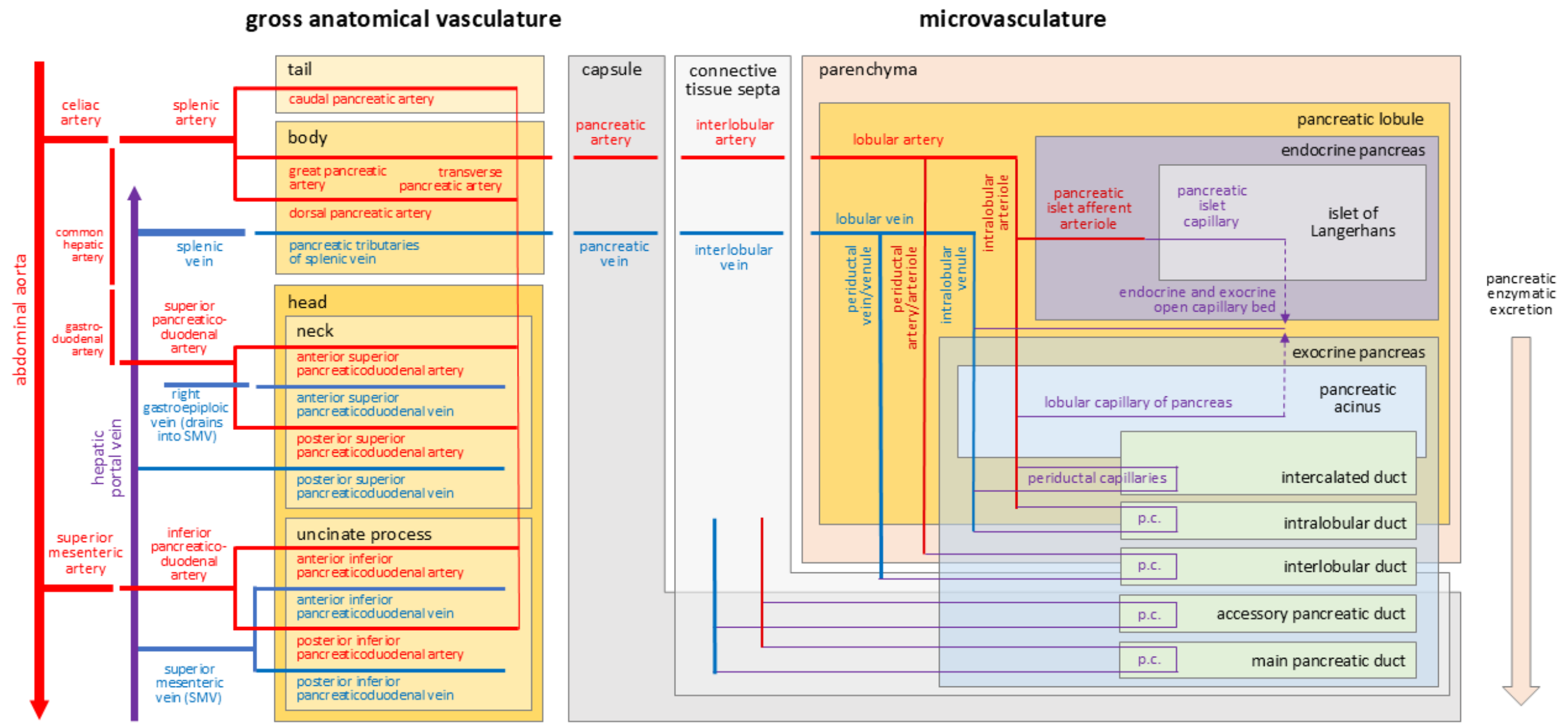

**Figure 3.** Blood vasculature pancreas crosswalk diagram.

## *Blood Vasculature Organ Crosswalk Table*

This initial release of the blood vasculature organ crosswalk table includes 349 vessels in 20 organs and body regions (**Table 1**). There are a total of 663 linkages from these vessels to individual anatomical structures. Blood vasculature organ crosswalk diagrams were created for 15 organs: brain, eye, heart, kidney, large intestine, liver, lung, mouth, pancreas, placenta, prostate, skin, small intestine, spleen, and thymus. Vessels in the pituitary gland are included in the diagram for the brain.

| Organ or Body Region | Blood Vessels | Anatomical Structures | Linkages | Crosswalk Diagram | Reviewer Initials |
|---|---|---|---|---|---|
| abdominal cavity | 4 | 3 | 5 | | KG |
| angiosome | 70 | 41 | 71 | | KG |
| brain | 20 | 14 | 25 | brain | QC |
| eye | 5 | 2 | 5 | eye | KG |
| face (mouth) | 17 | 16 | 19 | mouth | KMB, ARL |
| heart | 26 | 1 | 26 | heart | KG |
| kidney | 25 | 35 | 80 | kidney | SJ, MTV |
| large intestine | 1 | 1 | 1 | large intestine | KG |
| liver | 52 | 20 | 60 | liver | TK, AMM |
| lung | 52 | 46 | 157 | lung | GP |
| muscular system | 23 | 69 | 88 | | KG |
| pancreas | 36 | 18 | 60 | pancreas | MCT |
| pituitary gland | 6 | 2 | 8 | brain | KG |
| placenta | 1 | 1 | 1 | placenta | KG |
| prostate gland | 1 | 1 | 1 | prostate | KG |
| skeleton | 10 | 4 | 14 | | KG |
| skin of body | 2 | 2 | 3 | skin | KG |
| small intestine | 1 | 1 | 1 | small intestine | KG |
| spleen | 17 | 17 | 37 | spleen | BS |

**Table 1**. Summary of the blood vasculature organ crosswalk table by organ or body region.

The 663 linkages in the blood vasculature organ crosswalk table are divided into seven relationship types: 83 capillaries "directly_supplies_and_drains" a structure; larger vessels, such as arteries and veins, either indirectly "supplies" (325 linkages) or "drains" (68 linkages) structures; 117 vessels that are completely contained within a larger structure are "part_of" that structure; 68 vessels that pass through a structure "overlaps" that structure; and, as a special case, the glomerular capillary is "connected_to" the mesangium and "surrounded_by" the glomerular capsule.

The blood vasculature organ crosswalk table includes linkages to "angiosomes", which are 41 three-dimensional regions of skin, muscle, and bone supplied by a primary artery[12]. They are clinically important, especially in plastic surgery. The crosswalk table links blood vessels to each angiosome as a whole structure as well as to individual muscles and bones within those angiosomes.

The blood vasculature organ crosswalk crosswalk table also includes vessels linked to 28 functional tissue units in 11 organs (**Table 2**).

| **Organ** | **Functional Tissue Units (FTUs)** |
|---|---|
| face (mouth) | ● gingival attachment<br>● parotid gland ducto-acinar unit<br>● sublingual gland ducto-acinar unit<br>● submandibular gland ducto-acinar unit |
| kidney | ● cortical collecting duct<br>● descending limb of loop of Henle<br>● inner medullary collecting duct<br>● loop of Henle ascending limb thin segment<br>● nephron<br>● outer medullary collecting duct<br>● renal corpuscle |
| | • thick ascending limb of loop of Henle |
| large intestine | • crypt of Lieberkuhn |
| liver | • liver lobule |
| lung | • alveolus of lung<br>• bronchus submucosal gland |
| pancreas | • intercalated duct<br>• islet of Langerhans<br>• pancreatic acinus<br>• chorionic terminal villus |
| prostate gland | • prostate glandular acinus |
| skin of body | • dermal papilla<br>• epidermal ridge of digit<br>• hair follicle |
| small intestine | • intestinal villus |
| spleen | • red pulp of spleen<br>• white pulp of spleen |
| thymus | • thymus lobule |

**Table 2**. Functional tissue units by organ in the blood vasculature organ crosswalk table.

### *Pathway Crosswalk Table*

The pathway crosswalk table contains 204 links between blood vessels, lymphatic structures, and nerves. Of these, 28 use the relationships type "adjacent_to" to group neurovascular bundles consisting of an artery, vein, lymph vessel, and nerve. Another 38 use "connected_to" relationships to connect capillaries to arterioles, venules, or veins. The blood vasculature ASCT+B table is organized as separate arterial and vascular trees, which end at a terminal vessel, such as a capillary or venule. The "connected_to" links are needed to link these to form continuous vascular loops that extend from the heart to the organ and back to the heart. Another 138 "anastomosis_to" relationships indicate additional connections between vessels, which enable alternative routes for blood to flow.

### *Dataset Columns*

In the crosswalk tables, "ID" data fields link blood vessels and anatomical structures to standard ontologies. When possible, Uber anatomy ontology (UBERON) is used[13]. If there is no matching UBERON term, then the Foundational Model of Anatomy (FMA) ontology is used[14].

The data fields for the blood vasculature organ crosswalk table are

- **Vessel**. The name of the blood vessel.
- **VesselID**. The UBERON of FMA identifier for the vessel.

- **Relationship**. The type of relation from the blood vessel to the anatomical structure, such as “supplies” or “drains”.
- **BodySubPart**. The specific anatomical structure related to the blood vessel.
- **BodySubPartID**. The UBERON or FMA identifier for the BodySubPart.
- **BodySubPartSequence**. The order in which a vessel relates to multiple structures. A value of 999 means either the sequence order is unknown or will be added in a future version of this file.
- **BodyPart**. The organ or body region containing the BodySubPart.
- **BodyPartID**. The UBERON or FMA identifier for the BodyPart.
- **FTU**. If applicable, the name of the functional tissue unit.
- **FTULabel**. The UBERON or FMA label for the FTU.
- **FTUID**. The UBERON or FMA identifier for the FTU.
- **Angiosome**. If applicable, the name of the angiosome.
- **AngiosomeID**. The numeric identifier of the angiosome.

The data fields for the pathway crosswalk table are

- **Structure1Name**. The first blood vessel, lymph vessel, or nerve in the linkage.
- **Structure1ID**. The UBERON or FMA identifier for the first structure.
- **Structure1Type**. The first structure’s type (e.g., artery, lymph vessel, or nerve).
- **Relationship**. The kind of relationship between the two structures (e.g., “adjacent_to”, “connected_to”, or “anastomosis_to”).
- **Structure2Name**. The second blood vessel, lymph vessel, or nerve.
- **Structure2ID**. The UBERON or FMA identifier for the second structure.
- **Structure2Type**. The second structure’s type (e.g., artery, lymph vessel, or nerve).

# Technical Validation

## *Iteratively refining and validating the crosswalk diagrams and crosswalk tables*

The initial draft version of each blood vasculature organ crosswalk diagram was developed by an author of this study (K.G., Q.C., A.B., or G.W.) using domain knowledge and review of anatomy and histology textbooks[15–17]. The Radiopaedia.org website was also a useful resource, as well as two studies that provided extensive detail about the microvasculature of the brain cortex[18] and large intestine wall[19]. Linking blood vessels to angiosomes was informed by Taylor[12]. The draft was then reviewed by a second author (K.G or Q.C.) or one or more external organ experts (for kidney, liver, lung, mouth, pancreas, and spleen, see Acknowledgments).

This was an iterative process, which often took multiple rounds of review and refinement. The diagrams look quite different from typical anatomical illustrations of an organ. They are a highly distorted view of an organ designed to emphasize the relationships between the organ’s anatomical structures and the blood vessel pathways. They intentionally force the domain experts to focus on the organ's vasculature and not the anatomy or physiological function of the organ. It often requires trying several different layouts, before successfully "unraveling" the blood vessel pathways into simple loops.

However, by doing this, it becomes obvious where there are problems with the diagram or gaps in knowledge about parts of an organ's vasculature. They literally appear as gaps in the blood vessel loops, blood vessels that do not pass through any organ structure, or organ structures that do have blood vessels crossing them. In contrast, in a traditional illustration of an organ, the three-dimensional blood vessel pathways are difficult to trace, are hidden by other structures, or unclear. Smaller vessels and microvasculature, in particular, are often depicted as random branching lines to indicate their presence, but without being precise about their actual connections and paths.

### *Identifying gaps in knowledge using the diagrams and crosswalk tables*

An important outcome from these expert reviews was identifying parts of an organ where the vasculature is not yet well described. Often, the domain experts were not aware of these knowledge gaps before reviewing the diagram. They only realised it when the vasculature was presented this way and they could not fill in certain missing sections of the diagrams.

For example, gross anatomical scale vasculature (the main arteries that supply an organ) and microvascular (capillaries in functional tissue units) are typically viewed separately, either directly with the eye or through a microscope, respectively. However, details about the transition between the two through multiple branching levels of smaller and smaller unnamed arteries and arterioles was less understood. Though, these are explicitly drawn in the diagrams and a necessary part of the VCCF. There was similar uncertainty from domain experts in the venous drainage of organs and blood supply to supporting structures, such as collecting ducts.

### *Mapping to Standard Ontologies*

Of the 349 vessels in the blood vasculature organ crosswalk, 112 are mapped to UBERON IDs. For those vessels without an UBERON ID, 155 are mapped to FMA IDs. There is ongoing work within the HRA to generate new UBERON IDs for the remaining 82 vessels.

### *Algorithmic Validation*

In a prior study, we programmatically used the two crosswalk tables, in combination with other HRA data tables, to trace the full vascular pathways from the heart to each functional tissue unit and back to the heart. This validated that there were no gaps along these pathways in the datasets[9].

### *Limitations*

This work is far from complete. The HRA is an ongoing effort to map the entire human body. There are major releases of updated HRA digital objects every six months. The blood vasculature organ crosswalk and pathway crosswalk tables presented here reflect the state of the HRA as of its 11th release in June, 2026. These will continue to improve over time, not only as we expand the number of organs and structures in the crosswalks, but also as new knowledge about the human body is generated by the consortia that are directly contributing to the HRA and the broader scientific community.

The blood vasculature organ crosswalk diagrams are not final. Some are at a very early stage, include question marks ("?") for uncertain vessels, or have other comments in them. They are evolving documents that capture a snapshot in time of the HRA, and they will continue to be updated, but they might never be fully complete as the human vasculature is vast. Importantly, many vessels and linkages in the diagrams have not yet been recorded in the crosswalk tables. This may be because additional validation is still needed, or because certain vessels have been prioritized by the HRA initiative over others.

With each HRA release, the two crosswalk tables might become out-of-synch with other HRA digital objects. In subsequent HRA releases, we strive to resolve these issues.

## Usage Notes

The blood vasculature organ crosswalk and pathway crosswalk tables connect ASCT+B tables, which are available on the HRA portal at https://humanatlas.io. More information about the VCCF is available at https://humanatlas.io/vccf.

The HRA portal also includes links to standard operating procedures (SOPs, https://humanatlas.io/standard-operating-procedures), including the SOPs for authoring the blood vasculature organ crosswalk table (https://zenodo.org/records/11623745), the pathway crosswalk table (https://zenodo.org/records/11623415), and the blood vasculature organ crosswalk diagrams (https://zenodo.org/records/11623898).

To join the HRA initiative, sign-up for the open-to-all, monthly HRA Working Group meetings at

https://iu.co1.qualtrics.com/jfe/form/SV_bpaBhIr8XfdiNRH.

# Acknowledgments

We thank our organ experts, Sanjay Jain and M. Todd Valerius (kidney), Timothy Kendall and Anna Maria Masci (liver), Gloria Pryhuber (lung), Kevin Matthew Byrd and Angelo Rosso Llantada (mouth), Martha Campbell-Thompson (pancreas), and Birte Steiniger (spleen), for their feedback on the blood vasculature organ crosswalk diagrams.

We also thank Ellen M. Quardokus for connecting us with these experts; Aleix Puig Barbe for suggestions on ontologies for the crosswalks; Parth Kedar Godse, Zeynep Saziye Amac, and Supriya Bidanta for help with visualization and validation of the dataset; and Zorina Galis for the idea to create a common coordinate framework of the human body based on the vasculature.

The Human Reference Atlas (HRA) is under active development by the Indiana University Mapping Component as part of the HuBMAP HIVE, SenNet CODCC, KPMP, GUDMAP, and CIFAR efforts with expert input by the HRA Editorial Board and in close collaboration with experts from more than 25 other consortia. Data was provided by HuBMAP and other Tissue Mapping Centers. This research has been supported by the NIH Common Fund through the Office of Strategic Coordination/Office of the NIH Director under awards OT2OD033756 and OT2OD026671, by the Cellular Senescence Network (SenNet) Consortium through the Consortium Organization and Data Coordinating Center (CODCC) under award number U24CA268108, and by the NIDDK under awards U24DK135157 and U01DK133090. The work has also been supported by the Kidney Precision Medicine Project grant U2CDK114886, and the NIH National Institute of Allergy and Infectious Diseases (NIAID), Department of Health and Human Services under BCBB Support Services Contract HHSN316201300006W/HHSN27200002. K.B. is a co-director of and is funded by the CIFAR MacMillan Multiscale Human program. The content is solely the responsibility of the authors and does not necessarily represent the official views of the National Institutes of Health.

# Author contributions

K.G., Q.C., A.B., and G.W. developed the blood vasculature organ crosswalk diagrams and crosswalk tables; B.H. integrated the crosswalk tables with the HRA portal and knowledge graph; K.G., Q.C., and G.W. wrote the original manuscript draft; all authors reviewed and edited the manuscript; G.W. leads the VCCF; and, K.B. leads the overall HRA effort and obtained funding.

# Competing interests

G.W. is a paid consultant on the National Institutes of Health Human BioMolecular Atlas Program (HuBMAP) awards OT2OD033756 and OT2OD026671.